\documentstyle[11pt]{article}

\input{tcilatex}

\begin{document}

\title{Creation of Spin-1/2 Particles in the Hyperboloid de Sitter Space-Time}
\author{Ali Havare, Taylan Yetkin, Murat Korunur \\
Department of Physics, Mersin University,  33342, Mersin-Turkey}
\date{\today}
\maketitle

\begin{abstract}
In this work we solve Dirac equation by using the method of seperation of
variables. Then we analyzed the particle creation process. To compute the
density number of particles created Bogoliubov transformation technique is
used.
\end{abstract}

\section{Introduction}

The creation of elementary particles in curved space-time by strong fields
is one of the most exciting problem and still matter of investigation in
contemporary theoretical physics. In the literature a lot of studies can be
found concerning with the phenomenon of particle creation, mainly in
four-dimensional isotropic and homogeneous cosmological models as well as
two-dimensional toy models\cite{1,2,3,4}. In curved space-time quantum field
vacuum must be defined carefully since the vacuum plays an important role in
understanding the nature of particle creation. One of the most studied
technique to compute the densitiy number of particles created is the
Bogoliubov Transformation Technique(BTT)\cite{5}. In this technique `in' and
`out' vacuum states are determined from the solutions of relativistic wave
equations using a particular choice of time. Then the vacuum is specified $%
\left| in,0\right\rangle $ and $\left| out,0\right\rangle ,$ and the
Bogoliubov mixing transformation is found between them; 
\begin{equation}
\Psi _{in,k}(x)=\alpha \Psi _{out,k}(x)+\beta \Psi _{out,-k}^{\ast }(x),
\label{1}
\end{equation}
where $\alpha $ and $\beta $ are mixing coefficients, and they are related
to the density number of particles created. The studies in the literature
have considered four-dimensional isotropic space-times as well as
two-dimensional simple models.

In this work we will use two-dimensional de Sitter space-time represented as
the hyperboloid 
\begin{equation}
z_{0}^{2}-z_{1}^{2}-z_{2}^{2}=-a^{2}  \label{2}
\end{equation}
embeded in three dimensional Minkowski space\cite{6}, where 
\begin{eqnarray}
z_{0} &=&a\sinh (t/a),  \label{3} \\
z_{1} &=&a\cosh (t/a)\cos x,  \label{4} \\
z_{2} &=&\cosh (t/a)\sin x.  \label{5}
\end{eqnarray}
The line element of the space-time is given by 
\begin{equation}
ds^{2}=-dt^{2}+a^{2}\cosh ^{2}(t/a)dx^{2},(-\pi <x<\pi ).  \label{6}
\end{equation}
Because of the simple form of the line element the solution of the Dirac
equation and definition of Bogoliubov coefficients are easy. For the model
in Eq.(\ref{5}) the Hawking effect for scalar particles has been studied by
using BTT\cite{7}.

\section{Solution of the Dirac Equation}

The covariant generalization of Dirac equation is 
\begin{equation}
\left[ i\gamma ^{\mu }(\partial _{\mu }-\Gamma _{\mu })-m\right] \Psi (x)=0,
\label{7}
\end{equation}
where $\gamma ^{\mu }$ are curved space-time Dirac gamma mattricies related
to flat space-time gammas as $\gamma ^{\mu }(x)=e_{(i)}^{\mu }\gamma ^{(i)}$
with the tetrad definition  
\begin{equation}
e_{(i)}^{\mu }e_{(k)}^{\nu }\eta ^{(i)(k)}=g^{\mu \nu }.  \label{8}
\end{equation}
The spin connections $\Gamma _{\lambda }$ are defined 
\begin{equation}
\Gamma _{\lambda }=-\frac{1}{8}g_{\mu \alpha }\Gamma _{\nu \lambda }^{\alpha
}[\gamma ^{\mu },\gamma ^{\nu }],  \label{9}
\end{equation}
where $\Gamma _{\nu \lambda }^{\alpha }$ are Christoffel symbols given 
\begin{equation}
\Gamma _{\nu \lambda }^{\alpha }=\frac{1}{2}g^{\alpha \beta }(\partial _{\nu
}g_{\lambda \beta }+\partial _{\lambda }g_{\beta \lambda }-\partial _{\beta
}g_{\lambda \nu }).  \label{10}
\end{equation}
We can simplify calculations by introducing a conformal time given 
\begin{equation}
\eta =2\arctan [\exp (t/a)],  \label{11}
\end{equation}
then Eq.(\ref{6}) becomes 
\begin{equation}
ds^{2}=\frac{a^{2}}{\sin ^{2}\eta }(-d\eta ^{2}+dx^{2}),(0<\eta <\pi ).
\label{12}
\end{equation}
The Dirac equation Eq.(\ref{7}), after multiplying $-\gamma ^{(0)}$ from
left, becomes 
\begin{equation}
(\partial _{\eta }-\gamma ^{(0)}\gamma ^{(1)}ik-\frac{ima}{\sin \eta }\gamma
^{(0)})\Phi (\eta )=0,  \label{13}
\end{equation}
where the function of the form 
\begin{equation}
\Psi =\frac{1}{\sqrt{\sin x}}e^{ikx}\Phi (\eta )  \label{14}
\end{equation}
introduced to separate variables and to cancel out the terms coming from
spin connections. If we choose Dirac matrices in two dimensions as ($%
\{\gamma ^{(i)},\gamma ^{(k)}\}=2\eta ^{(i)(k)}$) 
\begin{equation}
\gamma ^{(0)}=i\sigma ^{3},\gamma ^{(1)}=\sigma ^{1},  \label{15}
\end{equation}
Eq.(\ref{13}) gives two coupled equations 
\begin{eqnarray}
(\partial _{\eta }+\frac{ma}{\sin \eta })\Phi _{1}(\eta )-k\Phi _{2}(\eta )
&=&0,  \label{16} \\
(\partial _{\eta }-\frac{ma}{\sin \eta })\Phi _{2}(\eta )+k\Phi _{1}(\eta )
&=&0,  \label{17}
\end{eqnarray}
It is easy to see that each component of the spinor satisfies hypergeometric
differential equation if we introduce the following functions in Eqs.(\ref
{16}) and (\ref{17}) 
\begin{eqnarray}
\Phi _{1}(\eta ) &=&\sin ^{ma}\eta \sin \frac{\eta }{2}f_{1}(\eta ),
\label{18} \\
\Phi _{2}(\eta ) &=&\sin ^{ma}\eta \cos \frac{\eta }{2}f_{2}(\eta ).
\label{19}
\end{eqnarray}
Then we obtain 
\begin{eqnarray}
\lbrack (x-1)\partial _{x}+\alpha ]f_{1}+kf_{2}, &=&0,  \label{20} \\
\lbrack (x+1)\partial _{x}+\alpha ]f_{2}+kf_{1}, &=&0,  \label{21}
\end{eqnarray}
where 
\begin{equation}
\alpha =ma+1/2;\,\,x=\cos \eta .  \label{22}
\end{equation}
Substituting Eq.(\ref{20}) into Eq.(\ref{21}) we obtain 
\begin{equation}
\left\{ u(1-u)\partial _{u}^{2}+[C-(A+B+1)u]\partial _{u}-AB\right\}
f_{1}(u)=0,  \label{23}
\end{equation}
where 
\begin{equation}
u=\frac{x+1}{2};A_{1,2}=ma+\frac{1}{2}\pm k;B_{1,2}=ma+\frac{1}{2}\mp k;C=ma+%
\frac{1}{2}.  \label{24}
\end{equation}
Solution to Eq.(\ref{23}) is\cite{8} 
\[
f_{1}(x)=N\,_{2}F_{1}(ma+\frac{1}{2}\pm k,ma+\frac{1}{2}\mp k,ma+\frac{1}{2};%
\frac{x+1}{2})
\]
\begin{equation}
+M\left( \frac{x+1}{2}\right) ^{\frac{1}{2}-ma}\,_{2}F_{1}(\pm k+1,\mp
k+1,-ma-\frac{3}{2};\frac{x+1}{2}),  \label{25}
\end{equation}
where $N$ and $M$ are normalization constants. The second component of the
spinor can be found using the recursion relations of the hypergeometric
functions. Then the exact solution of the first component of the spinor is
found as follow: 
\[
\Psi _{1}(\eta ,x)=e^{ikx}\sin ^{ma-\frac{1}{2}}\eta \sin \frac{\eta }{2}%
[N\,_{2}F_{1}(ma+\frac{1}{2}\pm k,ma+\frac{1}{2}\mp k,ma+\frac{1}{2};\frac{%
\cos \eta +1}{2})
\]
\begin{equation}
+M\left( \frac{\cos \eta +1}{2}\right) ^{\frac{1}{2}-ma}\,_{2}F_{1}(\pm
k+1,\mp k+1,-ma-\frac{3}{2};\frac{\cos \eta +1}{2})].  \label{26}
\end{equation}

\section{Particle Creation}

In order to analyze the phenomenon of the particle creation we proceed to
discuss the behaviour of the solutions of the Dirac equation in the
hypersurfaces $\eta =0,\pi $ or equivalently when $x=\pm 1.$ To relate two
different vacuum for $x=\pm 1$, we can make use of the relation for the
hypergeometric functions, 
\[
_{2}F_{1}(\alpha ,\beta ,\gamma ;z)=\frac{\Gamma (\gamma )\Gamma (\gamma
-\alpha -\beta )}{\Gamma (\gamma -\alpha )\Gamma (\gamma -\beta )}%
\,_{2}F_{1}(\alpha ,\beta ,\alpha +\beta -\gamma +1;1-z)
\]
\begin{equation}
+\frac{\Gamma (\gamma )\Gamma (\alpha +\beta -\gamma )}{\Gamma (\alpha
)\Gamma (\beta )}\,_{2}F_{1}(\gamma -\alpha ,\gamma -\beta ,\gamma -\alpha
-\beta +1;1-z),  \label{27}
\end{equation}
then the negative frequency mode for $x=-1$ reads 
\begin{equation}
f_{(x=-1)}^{-}=C^{-}\,_{2}F_{1}(ma+\frac{1}{2}\pm k,ma+\frac{1}{2}\mp k,ma+%
\frac{1}{2};\frac{x+1}{2})  \label{28}
\end{equation}
whereas the positive frequency mode for $x=+1\,$takes the form 
\begin{equation}
f_{(x=+1)}^{+}=C^{+}\,_{2}F_{1}(ma+\frac{1}{2}\pm k,ma+\frac{1}{2}\mp k,ma+%
\frac{1}{2};\frac{x+1}{2}),  \label{29}
\end{equation}
where $C^{+}$ and $C^{-}$ are normalization constants. Using the relation (%
\ref{36}) we obtain 
\begin{equation}
f_{(x=-1)}^{-}=\frac{\Gamma (ma+\frac{1}{2})\Gamma (-ma-\frac{1}{2})}{\Gamma
(k)\Gamma (-k)}\,f_{(x=+1)}^{+}+\frac{\Gamma (ma+\frac{1}{2})\Gamma (-ma-%
\frac{1}{2})}{\Gamma (ma+\frac{1}{2}+k)\Gamma (ma+\frac{1}{2}-k)}%
f_{(x=+1)}^{-}.  \label{30}
\end{equation}
Finally we obtain using Bogoliubov coefficients 
\begin{equation}
n\sim \frac{\left| \beta \right| ^{2}}{\left| \alpha \right| ^{2}}=\left| 
\frac{\Gamma (ma+\frac{1}{2}+k)\Gamma (ma+\frac{1}{2}-k)}{\Gamma (k)\Gamma
(-k)}\right| ^{2}  \label{31}
\end{equation}
From Eq.(\ref{31}) and relation $\left| \alpha \right| ^{2}+\left| \beta
\right| ^{2}=1$ we find that the density of particles created $\left| \beta
\right| ^{2}$ is given 
\begin{equation}
\left| \beta \right| ^{2}=\frac{\sin ^{2}k\pi }{\cos ^{2}(ma+\frac{1}{2}%
+k)\pi }.  \label{32}
\end{equation}
The expression (\ref{41}) shows that the density of particles created has a
maximum for $ma+k=n-\frac{1}{2}$, where n is integer. Also have that there
is no particle creation when $k$ takes integer values. Hence, if the
universe has the phenomenon of particle creation, the momentum can not take
integer values. When massless particles couples conformally to the metric 
\cite{3} Eq.(\ref{32}) takes the form 
\begin{equation}
\left| \beta \right| ^{2}=\frac{\sin ^{2}k\pi }{\cos ^{2}(\frac{1}{2}+k)\pi }%
=1.  \label{33}
\end{equation}
Equation (\ref{33}) shows that the density of particles created has a
maximum for negative modes.

For the de Sitter space-time studied by Villalba\cite{9} the density of
particles created is 
\begin{equation}
\left| \beta \right| ^{2}=\frac{1}{e^{2\pi m/H}+1}.  \label{34}
\end{equation}
We can see that both results in Eqs.(\ref{32}) and (\ref{34}) are depend on
the mass value of the particle while in Eq.(\ref{32}) it also depends on
momentum value of the particle.

\section{Conclusions}

We solved the Dirac equation using the merhod of seperation of variables for
de Sitter space-time. Then using the BTT method we computed that the density
number of particles created and also found that the particle creation does
not occur when the momentum of spin-1/2 particle has integer values.

\end{document}